**The control of graphene double-layer formation in copper-catalyzed chemical vapor deposition**

Martin Kalbac*, Otakar Frank and Ladislav Kavan

*J. Heyrovsky Institute of Physical Chemistry, Academy of Sciences of CR, Dolejskova 3, 18223 Prague 8, Czech Republic*

**Abstract**

The growth of graphene during Cu-catalyzed chemical vapor deposition was studied using $^{12}CH_4$ and $^{13}CH_4$ precursor gasses. We suggest that the growth begins by the formation of a multilayer cluster. This seed increases its size but the growth speed of a particular layer depends on its proximity to the copper surface. The layer closest to the substrate grows fastest and thus further limits the growth rate of the upper layers. Nevertheless, the growth of the upper layers continues until the copper surface is completely blocked. It is shown that the upper layers can be removed by modification of the conditions of the growth by hydrogen etching.

---

*Corresponding author. Fax: +420 28658 2307. E-mail address: martin.kalbac@jh-inst.cas.cz (M.Kalbac)

# 1. Introduction

The large scale production of graphene for electronic devices relies on catalytic chemical vapor deposition (CVD).[1] Therefore, main attention is dedicated to understand the mechanism of the graphene formation and also to control the growth. Nevertheless, in spite of many efforts put into the graphene CVD research, there are still many challenges to be solved. Cu or Ni are the most widely used catalysts due to their low cost, etchability and large grain size.[1,2,3,4] Depending on the catalyst, two mechanisms of the graphene growth are proposed.[5] In the case of Ni, the precursor is decomposed at the surface and carbon is dissolved in the metal. When the substrate is cooled down, the solubility of C in Ni decreases and graphene first segregates and then grows on Ni surface.[5,6] Hence, it is very important to control the cooling conditions to reach a monolayer graphene (1-LG).[7] On the other hand, in the case of copper catalyst, the carbon intermediate is not dissolved in the metal since the solubility of C in Cu is negligible even at a very high temperature. Instead, the carbon atoms form graphene directly on the surface already at high temperature, i.e. there is no need to precisely control the cooling of the metal. The CVD on copper is suggested to be surface mediated and self-limiting,[5] once the monolayer is completed, the process does not propagate any more, since the catalytic Cu surface is blocked. Hence, only 1-LG should be formed by the Cu-catalyzed CVD, but in many cases small regions with double- or multilayers are observed.[8] The mechanism of the formation of a multilayer regions is not well understood yet. These multilayer regions may impede the fabrication of graphene devices on large scale, because the multilayer areas disturb the uniformity of the graphene film.

Isotope engineering provides a unique possibility for advanced studies on graphene growth by Raman spectroscopy. In the presented study we use either $^{13}C$ or $^{12}C$ methane as the carbonaceous precursor to follow the Cu-catalyzed CVD synthesis of graphene in detail.

Based on these results we succeed to modify the growth conditions to suppress the presence of double-layers.

2. Experimental section

The graphene samples were synthesized using CVD as reported previously.[2,1] In brief: The Cu foil was heated to 1000 °C and annealed for 20 min under flowing $H_2$ (50 standard cubic centimeter per minute (sccm)). Then the foil was exposed to $^{13}CH_4$ for 3 min. and subsequently to $^{12}CH_4$ for 3 min. leaving hydrogen gas on with the same flow rate. Finally the substrate was cooled down quickly from 1000 °C to 500 °C under $H_2$. The etching of the top layers was realized by switching off the methane and leaving on the hydrogen gas for additional 1-20 min. at 1000°C. The pressure was kept at 0.35 Torr during the whole growth. The as-grown graphene was subsequently transferred to a clean $SiO_2$/Si substrate using polymethylmethacrylate (PMMA), according to procedures reported previously.[9] The Raman spectra were excited by 2.41 eV laser energy ($Ar^+$ laser, Coherent) and acquired by a LabRam HR spectrometer (Horiba Jobin-Yvon). The spectral resolution was about 1 $cm^{-1}$. The spectrometer was interfaced to a microscope (Olympus, 100x objective). Raman spectra were fitted by Lorentzian line shapes for the analysis. SEM imaging was performed using a high resolution SEM S-4800 (Hitachi).

3. Results and discussion

Figure 1 shows the optical image of graphene transferred onto a Si/$SiO_2$ substrate. The growth was, in this case, stopped before graphene fully covered the copper substrate in order to distinguish the individual graphene grains. An analysis of the optical contrast[10,11] in Figure 1 shows that the grains are formed from 1-LG except for the darker regions located in their centers. These spots correspond to double- or generally to multilayer graphene areas. Despite it is difficult to understand how the multilayers are formed at this stage, their almost exclusive centering on the monolayer grains probably excludes the possibility that the small grains are 'overgrown' by larger grains, which would grow faster or start to grow earlier. On the other hand, two other growth mechanisms may be envisaged. Yan et al.[12] proposed a sequential formation of the first layer, which might be eventually followed by the second one. We will discuss below an alternative mechanism of a simultaneous growth of all layers but with different speeds.

In order to understand the mechanism of the 2-LG formation in detail we modified our CVD and altered the precursor gas between $^{13}CH_4$ and $^{12}CH_4$, respectively during the synthesis. A similar approach has been already used to distinguish between precipitation growth mechanism and the surface growth mechanism on nickel and copper substrates.[13] If the $^{12}CH_4$ and $^{13}CH_4$ are altered, the resulting graphene grains are composed of either $^{12}C$ or $^{13}C$ labeled areas and the Raman spectroscopy can be used to distinguish them.

The frequency downshift of the Raman bands in the $^{13}C$ enriched material originates from the increased mass of this isotope according to equation (1): [13]

$$(\omega_0-\omega)/\omega_0 = 1 - [(12 + c_0)/(12+c)]^{1/2} \qquad (1)$$

where $\omega_0$ is the frequency of a particular Raman mode in the $^{12}C$ sample, $c = 0.99$ is the concentration of $^{13}C$ in the enriched sample, and $c_0=0.0107$ is the natural abundance of $^{13}C$. The frequencies of the G and the G' modes are about 1600 and 2700 cm$^{-1}$, respectively for the $^{12}C$ graphene. According to Eq. 1, the bands should redshift in $^{13}C$ graphene compared to $^{12}C$ graphene by about 100 and 200 cm$^{-1}$ for the G and the G' mode, respectively, The isotope shift is large enough to distinguish the corresponding Raman bands of the $^{12}C$ and $^{13}C$ graphene regions in various environments. We note that the frequency of the G and G' bands can be also influenced by stress[14,15] or local doping[16,17] but these effects are under normal conditions small compared to the isotope shift, therefore we do not consider them further in this study.

Figure 2a shows an optical image of a typical grain grown by alternating $^{13}CH_4$ and $^{12}CH_4$ gas sources. For easier identification of the number of layers the as-grown graphene was transferred from copper to Si/SiO$_2$ substrate. The grain in Figure 2 is formed by 1-LG except for a central area which is obviously 2-LG and can be easily identified by darker color. The graphene growth was started using $^{13}CH_4$, after 3 min the precursor gas was switched to $^{12}CH_4$ and the growth continued for subsequent 3 min. It should be emphasized that, compared to the previous work,[5] we switched $^{13}CH_4$ and $^{12}CH_4$ only once during the growth, hence we were able to clearly distinguish the different regions in a slowly growing layer. Furthermore, we limited a growth time to prevent merging of graphene grains. This simplifies identification of the multilayer spots within graphene grain and more importantly it also rules out the eventual formation of second-layer islands after the first layer completion as suggested previously.[12]

To analyze the growth, the Raman spectra were measured in profiles across the graphene grain. Typical results obtained when the Raman profile runs straight over both the 1- LG and 2-LG regions (i.e. across the centre of graphene grain) are presented in Figs. 2b-d and S1. As can be seen from the frequency of the Raman bands (Figs. 2d) the carbon isotope content is altered across the grain. Let us first focus on the intensities of the Raman G' and G bands in Figs. 2b and c, respectively, at the grain centre. Here, the Raman spectra consist only from contribution of $^{13}C$ graphene with intensities twice as high as in the rest of the grain (non-zero but very small contribution of the $^{12}C$ bands originates from the proximity of a $^{12}C$ region in the top layer, see below). The $^{13}CH_4$ was used for the first three minutes of the growth, hence the exclusive contribution of $^{13}C$ to the doubled intensity in the center of graphene grains suggests that the second layer is formed already at first stages of the growth. Therefore it can be suggested that middle 2-LG region of the grain actually corresponds to the initial graphene seed. Going slightly off the center of the graphene grain the contribution of $^{12}$C graphene in addition to the signal of $^{13}$C graphene starts to be apparent with their overall intensity still being at the level of the initial $^{13}$C bands. The signal of $^{12}$C graphene again disappears at the border of the central 2-LG region and, at the same time, the signal intensity of $^{13}$C is reduced by about 50% compared to the signal intensity in central 2-LG area. Going further in direction towards the grain edge, the signal of $^{12}$C graphene appears again, together with the $^{13}$C signal vanishing, and remains constant till the grain boundary. The presence of $^{12}$C in the double-layer region means that these regions grow during the whole growth, but obviously slower than the dominating single-layer part of the graphene grain. In other words, the top and bottom layers grow independently. We note that a spatial resolution of our spectrometer with 100x objective is about 0.5 µm, the size of the largest grains reaches typically about 20 µm after 6 minutes of the growth, hence we are able to distinguish about 9 s of the growth of the fast growing 1-LG at the given conditions. (Assuming for simplicity the speed of graphene growth is linear within this timeframe; see however, below.) On the other hand, the size of 2-LG central region is about 2 µm after 6 minutes of the growth, which gives the 'resolution' of 90 s and provides an estimate that the growth rate of the upper layer is smaller by a factor of ca. 10. This simple calculation shows an importance to switch between $^{13}$CH$_4$ and $^{12}$CH$_4$ only once during the growth to distinguish areas in 2-LG having different isotope composition.

From a practical point of view, it is also important to discuss whether the slowly growing layer is actually on top or underneath the dominating fast growing layer. As shown above, the slowly growing layer is formed simultaneously with the quickly growing one. Therefore in the case that the slowly growing layer is below the quickly growing layer, the precursor species would need to enter between the quickly growing layer and copper and then to travel a relatively large distance to reach the slowly growing layer. In addition, the slowly growing layer would need to lift off the quickly growing layer during the growth. Despite the relatively high growth temperature (1000 °C) we believe that these processes are unlikely. Therefore we suggest that the slowly growing layer is actually located on the top of the faster growing layer. Our suggestion is also in agreement with the recent work of Robertson and Warner.[18]

It was proposed previously that the chemisorption of methane on Cu with formation of $(CH_x)_s$ (x <4) surface-bound complex is thermodynamically unfavorable, but agglomeration into oligomeric $(C_nH_y)_s$ species is a thermodynamically favorable process ultimately leading to the growth of graphitic carbon.[19] Hence the formation of the multilayer graphene seed may be rationalized. The growth of layers depends on the distance between the particular graphene layer and the copper. Since methane splits at the top of the copper surface, the graphene layer closest to it will presumably exhibit the fastest growth. As the bottom layer spreads faster, the distance, which must be traveled by graphene precursors needed for the growth of the upper layers increases and the formation of these layers is slowed down even more. This finally leads to the relatively small total area of multilayer regions on single layer graphene. Also a more detailed analysis of the Figure 1 shows, that larger grains contain larger multilayer regions in the centre as compared to smaller grains. The smaller grains presumably begin to grow later but still contain a multilayer central region. Hence, this observation further supports the idea that the growth starts from a multilayer seed.

Within the experimental resolution mentioned above it seems that the growth speed of 2-LG region is similar for the $^{12}$C and $^{13}$C areas. However, the $^{13}$C area is grown during 3 min and within this time the size of the bottom quickly growing layer reaches already the size of about 10 µm. Hence, the change of the distance which must be traveled by carbon atoms to reach 2-LG region may not be relevant anymore. These results are also important for the understanding of the graphene growth mechanism. Assuming that the methane splits over Cu catalyst, the 'hot' graphene precursors should be able to travel around over relatively large distance to reach 2-LG in the centre of the graphene grain. This is in agreement with HR TEM results[20] where graphene samples heated by the electrical current showed reconstruction of edges due to moving of carbon atoms around a graphene sheet.

A practical consequence of the latter observation is that prolonging the time of the growth would also lead to an increased size of the multilayer regions if there is still a free copper surface. If the copper surface is fully covered by graphene, also the top layers stop to grow or this growth is almost suppressed since there are no carbonaceous intermediate precursors available.[21] Consequently, the formation of a complete double- or multilayer is unlikely.

In order to explore the size of the initial multilayer grains we further shortened the growth time to 5 s. The grains with a small size are more difficult to transfer and in addition the grains smaller than about 0.5 µm would not be possible to resolve by optical microscopy or micro Raman measurement. Therefore we visualized the grains directly on copper foil by scanning electron microscope (Figure 3). Despite a very short growth time, several graphene grains were found on the copper substrate. The final size of graphene grains after the growth was about 1 µm corresponding to a growth speed of about 12 µm/min. As mentioned above the six minutes growth leads to grain size of about 20 µm which corresponded to a growth speed of about 3.4 µm/min. This is still slightly faster than 0.5 µm /min reported previously for 10 min growth.[22] Hence it is obvious, that the growth speed is decreasing with the time of the growth especially at early stages of the growth. The previous study[22] suggested linear increase of the size with time of the growth but the first measurement considered in that study was after 2 minutes of the growth and at this time the growth rate might be already stable. As can be also seen from Figure 3, the middle region of the graphene grain is darker, which probably corresponds to a multilayer region. The multilayer region is relatively large with respect to the size of single layer region. Hence, our suggestion, that the layers grow simultaneously at first and in the later stages of the grain growth they decouple, is supported.

A straightforward approach to make graphene without multilayers would be the growth of one large domain which would suppress the formation of many multi-layer seeds. Despite a recent progress in the CVD graphene synthesis the domain size is limited to hundreds of micrometers, hence there is still a long way to reach a large scale production using such a method.[23] In addition, such procedures seem to require extremely clean conditions and long growth time[23] which would make the growth expensive. It was also suggested that the copper (111) leads to uniform monolayer.[24,25] Nevertheless, the use of single crystals for the graphene growth would be difficult to scale up.

Therefore we suggest a different approach. Since the slowly growing layers are located on top of the faster growing ones, they are well accessible for etchants. In addition, the slowly growing layers would have accessible reactive edges, which are more susceptible to the etching. If a multilayer-free 1-LG is required, it is thus possible to etch away the multilayers. In fact, such an etching step can be interfaced with the growth. Figure 4 shows an example of graphene prepared by a simple hydrogen etching, which followed immediately after the methane source was turned off while still maintaining both the hydrogen flow and also the temperature used for the growth. This shifts the equilibrium towards hydrocarbons formation and the multilayers are etched away. A similar approach has been used recently to reshape the graphene domains.[26] It was suggested previously, that hydrogen limits the growth[26] and it can etch the formed single layer graphene. We have not seen such an effect, because the etching step was realized after the growth when graphene completely covered the surface and no reactive sites on 1-LG were available. Nevertheless, it seems that non perfect areas –for example where two grains merge - can be also etched away if the hydrogen treatment is long enough. This is shown on Figure 4, where the growth of graphene was followed by prolonged hydrogen etching step. The 2-LG regions are clearly removed (cf. Figure 1) and in addition the contact areas between graphene grains are etched away. These results are also in agreement with previous studies of grain contact areas using HRTEM.[27] It was shown that there are defects and dangling bonds at the grain boundaries,[27] which leads to their increased reactivity, hence they should be preferentially etched by hydrogen.

## 4. Conclusion

In conclusion we present a study of the graphene growth using combination of $^{12}CH_4$ and $^{13}CH_4$ source gases. Our results suggest that the multilayer regions originate from the graphite-like seeds which are presumably formed at the beginning of the grain growth. Nevertheless, after initial stage of the growth, the top and bottom layers decouple and the top layer in average grows about 10 times slower than the bottom layer. The top layer is accessible for etchants and also it is reactive due to edge defects. Therefore it is possible to etch the top layer by hydrogen as we demonstrated by modified growth conditions.

**Figures and captions**

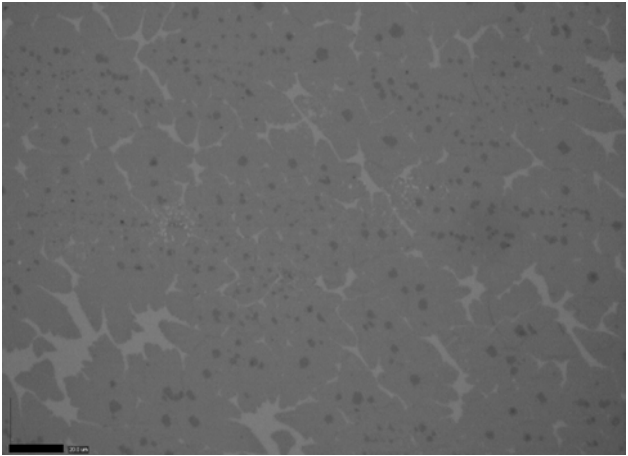

**Figure 1.** Optical microscope image of graphene grains with most of the area covered by 1-LG. The dark spots correspond to double-layer/multilayer regions in the center of each grain and the lightest regions correspond to a free SiO$_2$/Si surface. The scale bar is 20 µm.

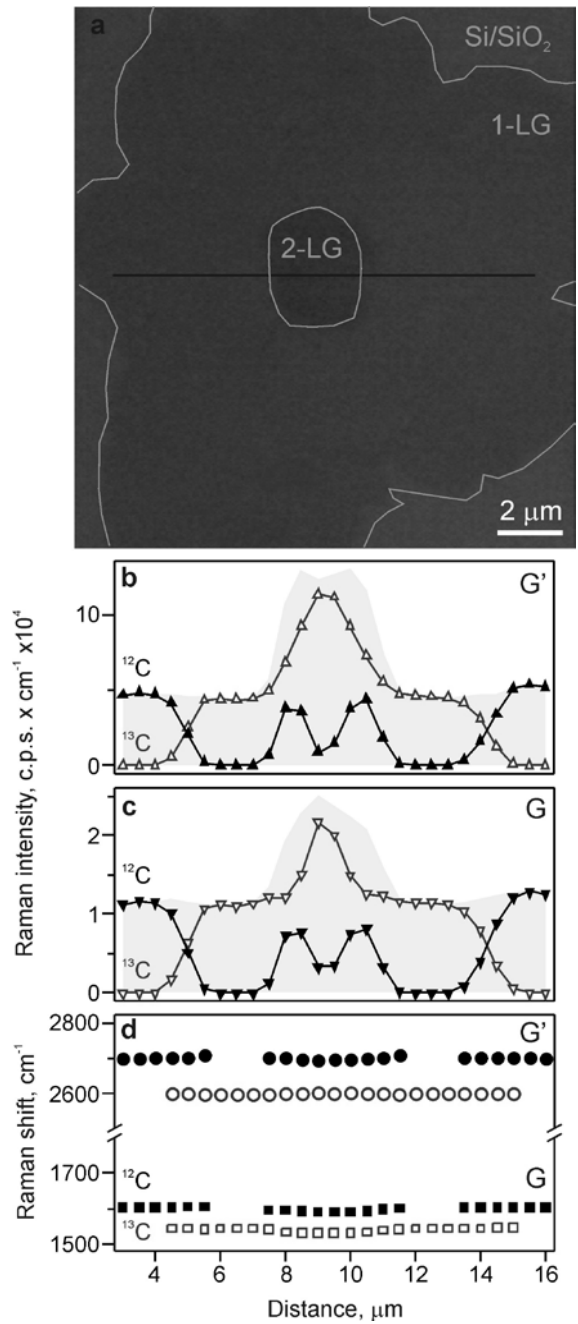

**Figure 2**. a) Optical microscope image of a 1-LG graphene grain on Si/SiO$_2$ substrate with 2-LG in the centre. b,c) intensity of the Raman G' and G bands as measured along the horizontal black line in a) with the grey-filled area representing the sum of the $^{12}$C and $^{13}$C bands for the particular mode. d) Raman shift of the G' and G bands along the profile.

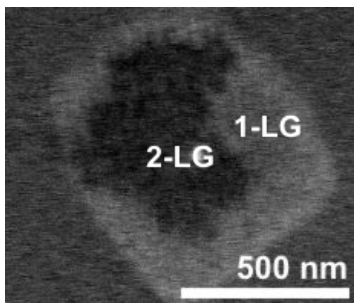

**Figure 3.** Scanning electron microscope image of a small graphene grain on a copper substrate grown during the initial 5 s of the reaction.

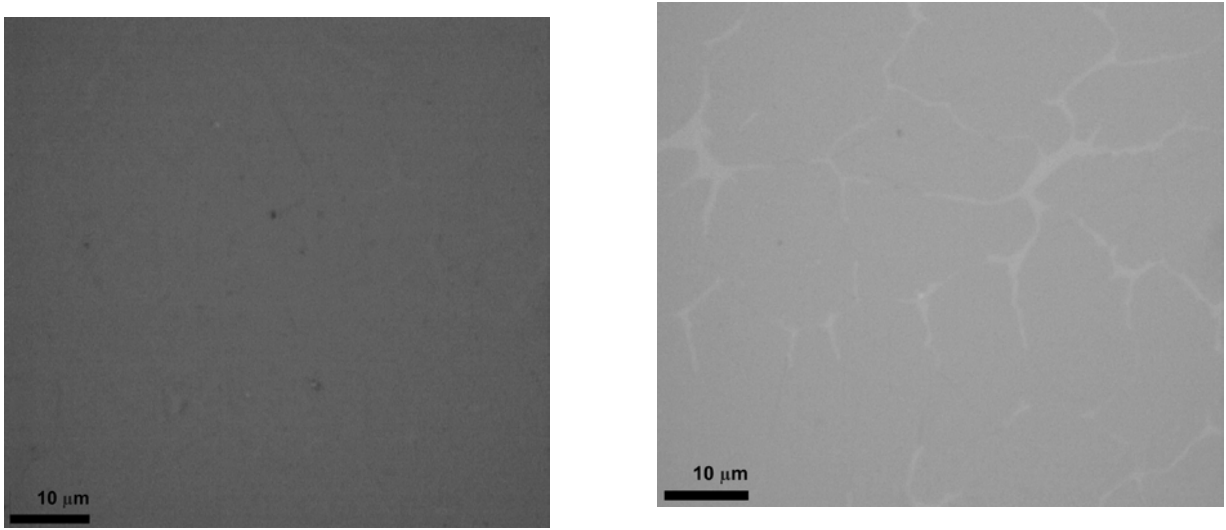

**Figure 4.** Optical microscope image of 1-LG graphene when the hydrogen etching step is included (left). Optical microscope image of 1-LG graphene grains after extensive hydrogen treatment, where the grain contact parts were etched away (right). The scale bar is 10 µm.

17